\documentclass[fleqn,usenatbib]{mnras}

\usepackage[T1]{fontenc}
\usepackage{ae,aecompl}
\usepackage{array,longtable}
\usepackage[ansinew]{inputenc}
\usepackage{lscape}
\usepackage{fancyhdr}

\usepackage{graphicx}	% Including figure files
\usepackage{amsmath}	% Advanced maths commands
\usepackage{amssymb}	% Extra maths symbols

\title[Spiral rotation speed of the Galaxy]
{The spiral pattern rotation speed of the Galaxy and the corotation radius with GAIA DR2}

\author[W. S. Dias et al.]{
W. S. Dias,$^{1}$\thanks{E-mail:wiltonsdias@yahoo.com.br}
H. Monteiro,$^{1}$,  J. R. D. L\'epine$^{2}$ ,  
and
D. A. Barros$^{3}$
\\
$^{1}$UNIFEI, Instituto de F\'isica e Qu\'imica, Universidade Federal de Itajub\'a, Av. BPS 1303 Pinheirinho, 37500-903 Itajub\'a, MG, Brazil\\
$^{2}${Universidade de S\~ao Paulo, Instituto de Astronomia, Geof\'isica e Ci\^encias Atmosf\'ericas, S\~ao Paulo - SP, Brazil}\\
$^3$Rua Sessenta e Tr\^es, 125, Rio Doce, Olinda, 53090-393 Pernambuco, Brazil
}

\date{Accepted 2019 April 24. Received 2019 April 22; in original form 2019 March 20}

\pubyear{2015}

\begin{document}
\label{firstpage}
\pagerange{\pageref{firstpage}--\pageref{lastpage}}
\maketitle

\begin{abstract}
In this work we revisit the issue of the rotation speed of the spiral arms and the location of the corotation radius of our Galaxy. This research was performed using homogeneous data set of young open clusters (age < 50 Myr) determined from Gaia DR2 data. The stellar astrometric membership were determined using proper motions and parallaxes, taking into account the full covariance matrix. The distance, age, reddening and metallicity of the clusters were determined by our non subjective multidimensional global optimization tool to fit theoretical isochrones to Gaia DR2 photometric data.  
The rotation speed of the arms is obtained from the relation between age and  angular distance of the birthplace of the clusters to the present-day position of the arms.
Using the clusters belonging to the Sagittarius-Carina, Local and Perseus arms, and adopting the Galactic parameters $R_0$ = 8.3 kpc and $V_0$ = 240 km\,s$^{-1}$, we determine
a pattern speed of $28.2 \pm 2.1$ km\,s$^{-1}$\,kpc$^{-1}$, with no difference between the arms.
 This implies that the corotation radius is $R_c = 8.51 \pm 0.64$ kpc, close to the solar Galactic orbit ($R_c/R_0 = 1.02\pm0.07$). 
\end{abstract}


%%%%%%%%%%%%%%%%% BODY OF PAPER %%%%%%%%%%%%%%%%%%

\section{Introduction}
The Gaia mission recently released its second set of data (DR2) \citep{GAIA-DR22018} which provides precise astrometric and photometric data for more than one billion stars with magnitude G less than 21. Among other scientific outcomes, a great improvement is expected in the understanding of the structure and dynamics of spiral arms of the Milky Way, thanks to the new data on spiral arms tracers. 
Among these, a major role will be played by the open clusters, due to their precise distance and ages that can be determined from the color magnitude diagrams.

The Milky Way is a grand design galaxy, as revealed by its large scale spiral structure (\citet{Georgelin1976A&A....49...57G}, \citet{Levine2006Sci...312.1773L}, \citet{Hou2009A&A...499..473H}, \citet{Hou2014A&A...569A.125H}, \citet{Reid2014ApJ...783..130R}). According to the classical theory (see \citet{Shu2016ARA&A..54..667S} for a review), the arms are located between the inner and outer Lindblad resonances. The radius of corotation ($Rc$), where the velocity of rotation of stars coincides with the rotation velocity of the spiral arms ($\Omega_p$), is a fundamental parameter, which divides the Galactic disk in two regions. The interstellar  matter penetrates the arms in opposite directions, and the arms change from leading to trailing at that radius. In one of the branches of the classical theory, the arms are small perturbations in the gravitational potential of the disk, like gravitational potential grooves or valleys, produced by the crowding of stellar orbits in some regions of the disk. The orbits of the stars constituting the arms are closed ones, which repeat themselves after each turn (in the rotating frame of reference of the arms), and this guarantees a relatively long lifetime to the spiral structure (\citet{Pichardo2003ApJ...582..230P},\citet{Junqueira2013A&A...550A..91J}).

In spite of its merits, the classical model is being seriously challenged. Another line of thinking considers that spiral arms are transient, appearing and disappearing in periods of time of a few hundred million years \citep{Sellwood2010MNRAS.409..145S}. In that view, $Rc$ is not an important reference, as it changes frequently. Others consider multiple spiral structures with distinct rotation speed (\citet{Quillen2005AJ....130..576Q} and \citet{Quillen2018MNRAS.480.3132Q}), which would mean that several $Rc$ coexist. 
Even among the researchers who adhere to the classical view of the spiral structure, there has been no consensus concerning the location of the $Rc$. \citet{Mishurov1999A&A...341...81M} and \citet{Dias2005} located it very close to the Solar orbit radius. However, \citet{Drimmel2001ApJ...556..181D} located it at 6.7 kpc, \citet{Acharova2012A&AT...27..359A} at 7 kpc, and \citet{Monguio2015A&A...577A.142M} places it farther away than the Perseus arm. 
Recently, \citet{Michtchenko2018ApJ...863L..37M} explained the moving groups observed in the solar neighborhood based on a model which places $Rc$ close to the Sun.

The lack of a consensus on $Rc$ shows the urgency of making use of the unprecedented volume and quality of Gaia DR2 data to establish firm observational constraints to it. The purpose of the present work is twofold. Firstly, we present a set of open clusters that has been analyzed using the Gaia DR2 catalogue. Secondly, we make use of this sample of open clusters to perform a new determination of $Rc$. We adopt a method similar to that of \citet{Dias2005}, which consists in integrating the orbits of the clusters back to their birthplaces. 
The clusters are believed to born in spiral arms, so that each cluster reveals the position of a spiral arm at a past time equal to its age. By comparing the past position of the arm with the present-day position, the rotation angle of the spiral structure during that interval of time is obtained. This method is straightforward and has the advantage that it does not require any model nor any unknown parameter, except for the rotation curve of the Galaxy. As will be discussed, the rotation curve is sufficiently well known. 

This paper is organized as follows: in the next Section we
describe the sample of open clusters. Section 3 presents the method of membership determination. In Section 4 we  comment on the method adopted for determining the distances and ages of the clusters. In Section 5 we discuss
the galactic rotation curves and the orbits of the open clusters. In Section 6
we present the approaches proposed to estimate the  rotation speed of the spiral pattern  and we present the results. We summarize the main conclusions and comment on their consequences in Section 7.

\section{The sample of young open clusters}

We used the version 3.5 of the DAML02 catalogue \citep{Dias2002} to select 441 clusters with ages lower than 50 Myr. In addition we also analyzed 154 open clusters with no age determined in DAML02, 31 clusters recently discovered by   \citet{Castro-Ginard2018arXiv180503045C} and 52 clusters discovered by \citet{Cantat-Gaudin2018arXiv180508726C}, looking for young open clusters. As detailed in Section 4, we performed the isochrone fit to the stars with membership greater than 0.51 in order to select the clusters younger than 50 Myr to compose our sample. 
%XX young open clusters with no previous age determination were incorporated into our sample.

For each cluster we searched for the stars in the Gaia DR2 catalogue, using the central coordinates and the radius taken from the DAML02 catalogue. Since the cluster's radius may be larger than that given in the catalogue (see \citet{vanLeeuwen2017}) we opted to use a region in the sky about 2 arcmin bigger than the area covered by the cluster. However, for the clusters with an apparent diameter less than 5 arcmin we searched for stars in an area 4 times the cluster area. 
In this way, we  include virtually all possible members of the studied clusters.

Before determining the membership, we filtered the data following the recipe published by \citet{Babusiaux2018}. After many tests we decided not to use the data of magnitude G. In this way we obtain the same values in the final results of the isochrone fit, but with smaller errors. In addition, the code  converges much faster, as discussed in detail by Monteiro and Dias (2019).  

The sample with a complete  vector (x, y, z, vx, vy, vz) of initial conditions to orbit integration is composed by 80 open clusters younger than 50 Myr. This final sample is relatively small mainly because of our choice to use only open clusters with parameters ($\mu_{\alpha}cos{\delta}$, $\mu_{\delta}$, radial velocity, distance, age) determined from the Gaia DR2 catalogue, and of the additional condition that the clusters had to be  approved in a visual inspection of the isochrone fit, as commented in the next section.

\section{Mean astrometric parameters and membership determination}
In a 3D plot of the Gaia DR2 astrometric data ($\mu_{\alpha}cos{\delta}$, $\mu_{\delta}$, $\varpi$) of the stars in the region of an open cluster, the members of the cluster appear in a clump. 
From the statistical point of view, the closer to the center of the clump is a star the greater the probability that it is a member of the cluster. So the membership problem can be formulated in the data space as a maximum likelihood one, based on the assumption of normally distributed uncertainties in proper motion and parallax as:   

\begin{equation}
f(\mathbf {X}) = {\frac {\exp \left(-{\frac {1}{2}}({\mathbf {X} }-{\boldsymbol {\mu }})^{\mathrm {T} }{\boldsymbol {\Sigma }}^{-1}({\mathbf {X} }-{\boldsymbol {\mu }})\right)}{\sqrt {(2\pi )^{k}|{\boldsymbol {\Sigma }}|}}}
\label{eq:multinormal}
\end{equation}
where $\mathbf {X}$ is the column vector ($\mu_{\alpha}cos{\delta}$, $\mu_{\delta}$, $\varpi$) composed of the proper motion components and the parallax, $|{\boldsymbol {\Sigma }}|\equiv \operatorname {det} {\boldsymbol {\Sigma }}$ is the determinant of the covariance matrix $\boldsymbol {\Sigma }$ and $\boldsymbol {\mu }$ the mean column vector. The covariance matrix incorporates all uncertainties and their correlations, which are all given in the Gaia DR2 catalogue.

Because of the large volume of data in Gaia DR2 and the large number of clusters to be analyzed, the resolution of the equations above can take a considerable amount of computational time due to the matrix operations involved. So, we opted to estimate the membership in two steps, as performed in \citet{Dias2018B2018MNRAS.481.3887D} and  Monteiro and Dias (2019). First we use the proper motion following the procedure described in \citet{Dias2018UCAC5} to select the stars with membership greater than 0.50 based on that model. Then we fit a 1D parallax space assuming Gaussian distributions for the field as well as for the cluster stars. 
With the distribution parameters estimated (mean proper motion and parallax) used as initial parameters we applied Eq. \ref{eq:multinormal} and the full variance and covariance data from Gaia DR2 to obtain final membership probabilities.

The solution of Equation \ref{eq:multinormal} provides the membership probabilities and the mean and standard deviation of the proper motions and parallaxes for each analyzed cluster. 

Finally, we used the stars with membership probability greater than 0.50 
and individual radial velocities given in Gaia DR2 to estimate the mean radial velocities of the clusters. In this study we opted to use the mean and standard deviation ($1\sigma$) to represent the radial velocity of the cluster. 
As we use only the clusters with parameters estimated using Gaia DR2 data to compose a homogeneous sample, this was a limiting criterion. In the final sample of 80 open clusters, 15 clusters have their radial velocity determined for the first time.

In Table \ref{tab:astrometric} we present the mean proper motion, mean parallax and mean radial velocity  with the respective errors represented by the one standard deviation of the open clusters provided by the
Eq. \ref{eq:multinormal}. In the table are also given the number of cluster members, the equatorial coordinates ($\alpha, \delta$) and the radius of each cluster.

\section{Distances and ages from the isochrone fit}
The next step in our procedure was to perform the isochrone fit by considering the astrometric membership of the stars in the cluster's field. The main objective was to determine the parameters: distance, age, color-excess and metallicity. The method to obtain distances and ages by isochrone fitting is described in previous papers of our group, applied to UBVRI data \citep{Caetano2015} and \citep{Monteiro2017} and more recently, applied to Gaia DR2 data \citep{Dias2018B2018MNRAS.481.3887D} and Monteiro and Dias (2019). We used exactly the same procedures in this work.

The parameters distance and age were obtained by fitting Padova theoretical isochrones \citep{Bressan2012} to clusters, applying the cross-entropy continuous multi-extremal optimization method, which takes into account Gaia DR2 photometric data guided by the astrometric membership.
Briefly, the main goal of the cross-entropy is to find a set of parameters for which the model provides the best description of the data as per maximum likelihood definition. This is performed by randomly generating N independent sets of model parameters and minimizing the objective function used to transmit the quality of the fit during the run process. 
To estimate the final errors on fundamental parameters, we used  Monte-Carlo technique, re-sampling in each run with a replacement
in the original data set, to perform a bootstrap procedure. The isochrones are also re-generated in each run from the adopted IMF.
The final fundamental parameters and the errors were estimated by the mean and one standard deviation of five runs.

In this work we opted to be very restrictive considering only the isochrone fit results which satisfy a visual inspection. Although this is not an objective criterion, the double check allowed us to debug the sample to include only the very best final results in the sample used in spiral rotation speed analysis. 

For a number of open clusters of the final sample, we determined values of distance and age which are in disagreement with published ones. A detailed discussion of each case is unnecessary, since  \citet{Dias2018B2018MNRAS.481.3887D} and the results obtained for the control sample from \citet{Moitinho2001}
show that the procedures and results obtained with Gaia DR2 data are reliable. In addition, the CMDs with our fitted isochrones were clearly better than those from the literature. A more detailed discussion will be presented in a forthcoming paper dedicated to the new open cluster catalogue based on Gaia DR2 data. 

In summary, our analysis of the rotation velocity of the arms and $Rc$ is based on new homogeneous results obtained exclusively with Gaia DR2 data and with our methods developed to estimate astrometric membership and mean parameters, as well as distances and ages from non-subjective isochrone fit.

The final fit results obtained for each cluster are presented in Table \ref{tab:photometric}\footnote{The results of the parameters of the clusters as well all the detailed plots of the isochrone fits are given in electronic format available at \url{https://wilton.unifei.edu.br/gaia-dr2/OCgaia.html}.}.

\section{The orbits of the open clusters and the Galactic rotation curve}

The gravitational potential of the Galaxy can be explained  by  a model  composed of an axisymmetric disk, a bulge, and a halo, with parameters adjusted in order to fit the rotation curve \citep{Barros2016A&A...593A.108B}. In practice,  the potential that provides the force field in the Galactic midplane, can be derived directly from the rotation curve, avoiding any discussion on the details of the model. 

Results based on recent important surveys show that the  rotation curve of the Galaxy is flat in the galactocentric radius interval of interest for this work,  between  about 6.5 kpc and 10.5 kpc. Since the orbits of the open clusters studied here are confined within this range, the shape of the rotation curve outside this interval has no influence. As examples of recently measured curves, see the one derived from maser sources \citep{2017AstBu..72..122R}, that derived from the SEGUE and RAVE surveys \citep{2018A&A...614A..63S}, and from GAIA DR2 \citep{2018arXiv181004445C}.
The new results are not discrepant with the older ones, like the HI + CO rotation curve of \citet{1989ApJ...342..272F} and those determined from high mass star forming regions with measured
parallax and proper motion by \citet{Reid2014ApJ...783..130R}. For simplicity we adopted a flat rotation curve which has been fitted to different tracers by \citet{Barros2016A&A...593A.108B}, for which we have an analytic expression.  
All the rotation curves have to be normalized to the adopted rotation velocity of the LSR ($V_0$) and to the solar radius ($R_0$), since better values of these parameters became available. We selected the most recent results from the literature; the value $R_{0} = 8.3$ kpc and $V_{0}=240$\,km s$^{-1}$ were taken from \citet{Gillessen-R0-2017} and \citet{Reid-V0-2016}, respectively.

\begin{figure}
\includegraphics[scale = 0.52]{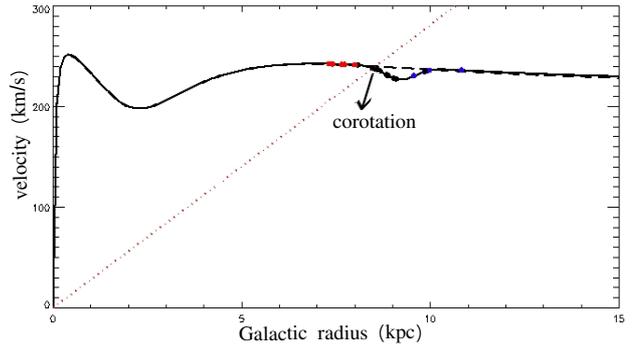}
\caption{Rotation curves of the Galaxy which can be classified as plane and plane with a dip centered at 9.2 kpc. The functions of the curves are presented in the equations \ref{eq:Vrot10} to \ref{eq:Vrot12}. The range of the Galactic distance of the sample of the open clusters and a respective arm is indicated by dots with color blue to Perseus, black to Local and red to Sagitarius-Carina. The dotted line indicates the spiral pattern rotation speed: $\Omega_p =28.2 \pm 2.1$ km\,s$^{-1}$\,kpc$^{-1}$.}
  \label{fig:curvaderotacao}
\end{figure}

There is, however, a small departure from flatness, in the form of a narrow dip at about 9.2 kpc, on which there is not yet a consensus. This feature  can be seen in the curves published by several authors , eg. \citet{1989ApJ...342..272F}, but has most often been ignored. It was tentatively explained by \citet{2009PASJ...61..227S} and by \citet{Barros2016A&A...593A.108B} as being due to 
a ring of matter in the galactic disk.  It cannot be excluded that the mass excess beyond the solar radius to be due to the Local arm, in which case it would only correspond to a sector of a ring, not extending along the whole galactocentric circle. Anyway, the important fact is that the dip does exist; in the absence of certainty on its nature, we adopt the description of the dip given by \citet{Barros2016A&A...593A.108B}, and more recently by \citet{Michtchenko2017A&A...597A..39M},
now slightly modified because of the new $R_0$ adopted here.
We used two rotation curves to determine $\Omega_p$ and  the location of $R_{c}$. One curve is flat close to the Sun (equation \ref{eq:Vrot10}), and the other is flat with a narrow dip (width 0.39 kpc, depth 12.5 km\,s$^{-1}$) centered at 9.2 kpc (equation \ref{eq:Vrot12}). Figure \ref{fig:curvaderotacao} presents the two curves, described by equations \ref{eq:Vrot10} and \ref{eq:Vrot12}.

%\begin{eqnarray}
%\label{eq:Vrot10}
% V_{\rm rot}(R) &=& %303\,\exp\left(-\frac&{R}{4.7}-\frac{0.036}&&{R}\right)  &\nonumber\\
% \qquad + &232\,\exp\left[-\frac&{R}{1400}-\left(\frac&{3.72}{R}\righ&% t)^2\right],\nonumber \\
%\end{eqnarray}

\begin {equation} \label{eq:Vrot10}
\begin{split}
V_{\rm rot}(R) &= &303\,\exp[-\frac{R}{4.7}-\frac{0.036}{R}] \\
& \qquad &+ &232\,\exp[-\frac{R}{1400}-(\frac{3.72}{R})^2]
\end{split}
\end{equation}

\begin{eqnarray}
\label{eq:Vrot12}
 V_{\rm rot}(R) &=& Eq.2 -12.5\,\exp[-\frac{1}{2}(\frac{R-9.2}{0.39})^2] \\ \nonumber
\end{eqnarray}

The present-day positions of the open clusters in the Galactic midplane, using a Cartesian coordinate system defined as follows, are obtained from the Galactic longitudes ($l$) and latitudes ($b$), derived from the equatorial coordinates $\alpha$ and $\delta$, and the heliocentric distances.

In all the figures of this paper we used the Galactic Center at (0,0) as reference and the x-axis pointing to the Galactic rotation direction. The rotation is clockwise or the vector angular velocity is perpendicular to the x-y plane pointing in the direction of the paper. The y-axis positive points towards the Galactic anti-center and the Sun is situated at (0,8.3) kpc position. 

To compute the heliocentric $U$ and $V$ velocities and the respective errors of each open cluster, we use the equatorial coordinates, distances, proper motions, and radial line-of-sight velocities, following the formalism described by \citet{Johnson&Soderblom1987AJ.....93..864J}, with $U$ positive towards the Galactic
anti-center and $V$ positive towards the direction of Galactic rotation. In order to pass to the local standard of rest (LSR) reference frame, we add the components of the solar motion $U_{\odot} = (-11.10\pm0.75)$\,km s$^{-1}$ and $V_{\odot} = (12.24\pm0.47)$\,km s$^{-1}$ \citep{Schonrich2010MNRAS.403.1829S}. 
The errors of $U_{LSR}$,$V_{LSR}$ velocities were determined by the usual propagation formula.

With the $U_{LSR}$ and $V_{LSR}$ velocities and the azimuths $\theta$, and adopting the LSR velocity
$V_{0}=240$\,km s$^{-1}$ \citep{Reid-V0-2016}, we calculate the Galactocentric radial $V_R$ and tangential $V_{\theta}$ velocities. 
For each open cluster, a vector with components $(R,\theta,\,V_{R},\,V_{\theta})$ is used as initial conditions for the integration of the orbit. 

The birthplace positions and uncertainties of
the open clusters were determined following \citet{Dinescu1999AJ....117.1792D}. 
We considered Gaussian distributions for the radial line-of-sight velocities,
proper motions, and heliocentric distances. The standard deviations
of the distributions were taken as being the measured errors of 
the observables. The birthplace positions were determined in a
Monte Carlo scheme, by integrating the orbits 1000 times from the 
generated Gaussian distributions. The birthplaces of the clusters
were taken as the mean of the distributions of the results of the
integrations, with the uncertainties represented by the standard
deviations. As in \citet{Dinescu1999AJ....117.1792D}, the solar Galactic radius,
the solar motion, and the velocity of the LSR were kept fixed in 
the integrations.

The numerical integrations of the equations of motion were performed by means of a fifth-order Runge-Kutta integration procedure,
with a typical time step of 0.1 Myr. The orbits of the open clusters were integrated backwards, for a time interval equal to the age of each object. This procedure provided the initial positions $X_0$ and $Y_0$, i.e. the birthplaces of the open clusters in the Galactic midplane. 

Table \ref{tab:galactic} gives the present-day (X,Y) and the birthplace ($X_0$,$Y_0$) positions of the open clusters in the Galactic midplane. The errors in the present-day ($\sigma X,\sigma Y$) positions were obtained by the usual propagation formula while the errors in the birthplace positions were determined by Monte Carlo procedure as explained earlier.

\section{Method, results and discussions}
The determination of the rotation speed of the spiral arms is based on the hypothesis that the birth of open clusters takes place in spiral arms. This follows from the ideas of \citet{Roberts1969ApJ...158..123R}, \citet{Shu1972ApJ...173..557S} and many others, according to whom the shock waves occurring in spiral arms are the triggering mechanism of
star formation. 

Therefore, every cluster carries with itself the evidence that a spiral arm was present at a given place (the birthplace) at a given epoch, specified by the age of the cluster. We find the birthplace of each cluster, integrating backward its orbit for time interval T equal to its age, starting from the present-day positions and space velocities, as discussed in the last section.
  
Basically, the birthplace of a cluster is supposed to represent a point of a spiral arm, a time T ago. So, if we rotate forward this point by an angle  $\Omega_pT$ around the Galactic Center, we obtain a point situated on the present-day position of the arm. 
Therefore, for each birthplace of a cluster, we measure the galactocentric rotation angle that should be applied to make it coincide with the corresponding present-day spiral arm.
Note that what is being rotated in this way back is a point of a spiral arm, and not the cluster.
In this way, clusters with  different ages produce independent measures of the rotation 
velocity, according to the equation \ref{kinematics} where the unknown parameter is $\Omega_p$. 

\begin{equation}
\theta_f = \theta_i + \Omega_{p}\times T
\label{kinematics}
\end{equation}
where, in polar coordinates, the azimuth $\theta_f$ is the present-day position angle of the arm, $\theta_i$ the birth-place position angle of a cluster, $\Omega_p$  the rotation velocity of the arms, and T is the age of the cluster. We consider that the shape of the spiral arms is conserved, so that every point of the initial spiral arm has a corresponding point on the present-day arm obtained by a rotation angle $\Delta\theta$ = $\theta_f$ - $\theta_i$. The concept of present-day spiral arm is next discussed.

\begin{figure}
\includegraphics[scale = 0.8]{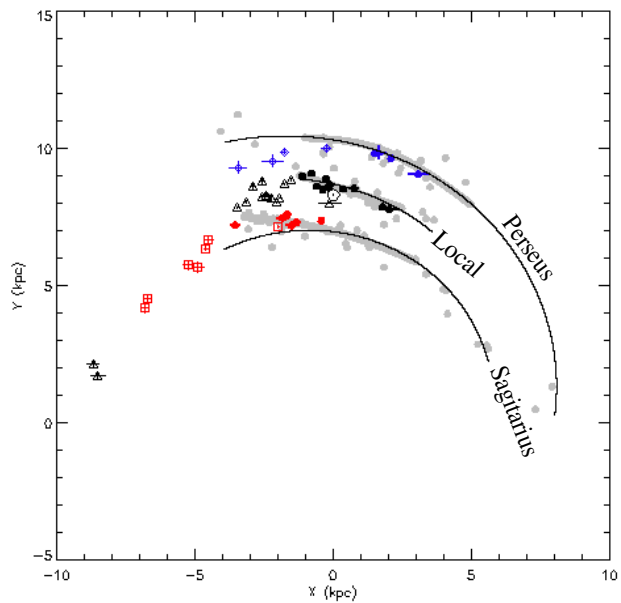}
\caption{Location of the present-day positions (dots circles) and birthplace (open symbols) of the open clusters of the sample. In light-gray are the polynomials fit for present zero-age arms positions traced by masers and HII regions. The small range of each arm are restricted to the range of interest limited by the positions of the open clusters used in this study. The black line curves are log-periodic functions from \citet{Reid2014ApJ...783..130R}. } 
\label{fig:atualenascimento}
\end{figure}

We adopted the present zero-age arms positions traced by masers sources since they are associated with massive O-B stars with lifetimes of the order of a Myr, which is small compared to the ages in our sample of clusters. We used the positions obtained from VLBI observations provided by \citet{Reid2014ApJ...783..130R}. Due to the lack of maser sources
on the Sagittarius-Carina arm on its extension to $x \leq -1$ kpc, we completed the sample with HII regions data with photometric distances or distances from parallax given by \citet{Hou2014A&A...569A.125H}.
As presented in Fig. \ref{fig:atualenascimento} by the light-gray points and curves and also in the Figure 1 of \citet{Hou2014A&A...569A.125H} the sample of masers and HII regions trace clearly the segment of the arms at present-day in the solar neighborhood.

We fitted polynomials with different degrees for each arm (degree 2 for Sagittarius-Carina, 1 for Local, 6 for Perseus) to the position of the masers, to obtain analytical description of the zero-age arms, in the regions of interest (the regions where we perform comparisons of the position of our sample of clusters with the arms). The reason for using  polynomials instead of logarithmic spirals is that the arms are known to present local deviations with respect to the global logarithmic spirals, see eg. the case of the Sagittarius-Carina arm  \citep{1999AstL...25..591B, 2006Sci...311...54X}.
In the case of present work, the aim is not to obtain a satisfactory fit over a large region of the Galaxy, but the best fit possible over the small region of interest. The polynomial fit gives more liberty to the adjusted curve to run along the small deviations from spirals. 

In Fig. \ref{fig:atualenascimento} are presented the birthplace and the present-day Galactic position of the clusters of the sample as well the present-day position of the arms Sagittarius-Carina, Local and Perseus.   
It shows clearly that the young open clusters are tracers of the spiral structure, as has been known and used by many authors for different goals (e.g., \citet{Becker1970, Russeil2003}).  
The present-day locations of the clusters very closely follow the present-day locations of spiral arms based on masers and HII regions. This requires that the clusters have not moved out of the spiral arms over their lifetimes, which implies the cluster orbital motions must be close to that of the arm pattern. Indeed, looking at Fig. \ref{fig:atualenascimento}, one might even be able to see that the Sagittarius arm clusters slightly lead the arm, the Local arm clusters are pretty well centered, and the Perseus arm sources slightly tend to lag the arm. Such a pattern would be expected for corotation near the radius of the Local arm.

\begin{figure*}
\includegraphics[scale = 0.42]{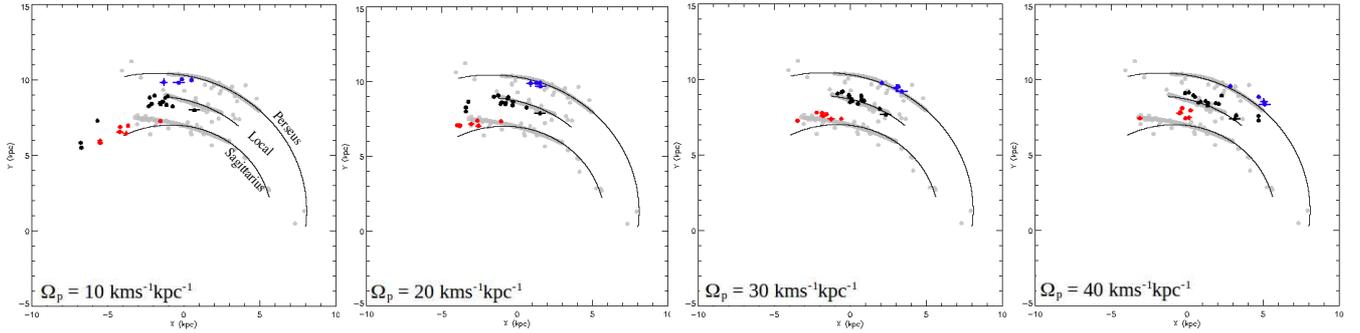}
\caption{Present zero-age arms positions traced by masers and HII regions. The color dots are the returned positions of the arms reconstituted by $\Omega_pT$ from the birthplaces of the clusters, using different values of $\Omega_p$.
See Fig. \ref{fig:atualenascimento} caption for other details. The final value of $\Omega_p$ is presented in the text.}
  \label{fig:diferentes-omegap}
\end{figure*}

Note that only the correct value of $\Omega_p$ produces rotated positions coincident with the zero-age positions of the arms as can be seen in Fig. \ref{fig:diferentes-omegap}.
The width of the arms used to test this coincidence is based on the work of \citet{Reid2014ApJ...783..130R}  which provides the intrinsic arm width using 103 regions of high-mass star formation measured with VLBI techniques. However, to accommodate the different values estimated for each arm as well as the uncertainties associated with the present-day and birth position of the open clusters we adopted a width of $\pm$ 0.3 kpc. If the rotated back position of a cluster did not fall within this distance from the zero-age arm, this cluster was rejected. This occurrence depends on the adopted value of $\Omega_p$. A fast converging process showed the value of $\Omega_p$ to be used. After calculating by the line fitting procedure described below, we 
entered the new $\Omega_p$ in the process above.

\begin{figure}
\includegraphics[scale= 0.8]{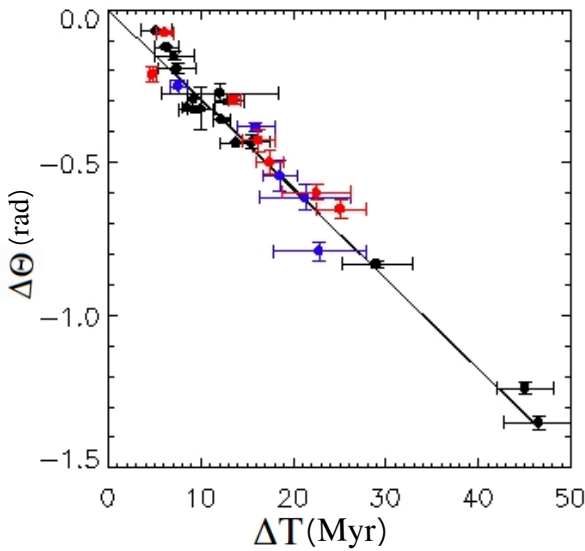}
\caption{Weighted linear least squares fit of $\Delta\theta$ vs. $\Delta T$. 
The slope gives the value of $\Omega_p = 28.2\pm2.1$ km\,s$^{-1}$\,kpc$^{-1}$.} 

  \label{fig:linearfit}
\end{figure}

Considering the zero-age arms as reference for the measurements of the displacement of birthplaces of clusters we applied the following method to estimate $\Omega_p$. We measure the angle $\Delta\theta$ that each cluster's birth position has to be rotated to coincide with the zero-age arm, as given in the equation \ref{kinematics}. In Fig. \ref{fig:linearfit} we present the plot with $\Delta\theta$ vs. the age of the cluster ($\Delta T$).

The weighted linear least squares fit of $\Delta\theta$ vs. $\Delta T$ for each arm gives  $\Omega_p$, which is the slope of the fit. The best fit is given by minimizing the chi-square error statistic for a one parameter linear function, i. e., just the angular coefficient. The weights are given by the uncertainty associated with each measurement.   
Tab. \ref{tab:omegap} presents the results for $\Omega_p$ for each arm. 

In Tab. \ref{tab:omegap} are also given the values of $\Omega_p$ obtained using the present-day arms described by spiral arms log-periodic functions as presented in Fig \ref{fig:atualenascimento}. To accommodate the uncertainty in the present-day spiral arms we opted to present the mean values determined considering polynomials and spiral arms log-periodic functions to describe the present-day arms local. The errors are obtained by combining quadratically the errors from each fit. 
In our opinion the polynomial and logarithmic spirals fits are in agreement in the region of interest of our work, which can be confirmed by the values of the $\Omega_p$. The difference found for the Sagitarium arm is due to the difference between polynomials and spiral for $X<0$, due to the lack of maser data, where the open clusters are concentrated.

We also tested the dependence of the results with the age of the clusters of the sample performing the same procedures using sub-samples with different range in age. The $\Omega_p$ obtained from the complete sample were reproduced within the errors. This provides an observational evidence that arms do not have any angular acceleration at least in the last 50 Myr.

The derived corotation radius ($Rc$) obtained by $Rc = \frac{Vrot(Rc)}{\Omega_p}$ is $R_c/R_0 = 1.02\pm0.07$ from both rotation curves. 
In the Fig. \ref{fig:curvaderotacao} we present the $\Omega_p$ line over plotted to the Galactic rotation curves used in this study as commented in details in the last Section. The intersection of the curves indicates the $Rc$ at about 8.5 kpc close the Galactic position of the orbit of the Sun. We
neglect the fact that the $\Omega_p$ line crosses the dip at a rotation velocity smaller than 240 $kms^{-1}$,
since the difference in velocity is very small.

It is interesting to point out we found an excellent agreement with the value of $R_c/R_0 = 1.06\pm0.08$ obtained in \citet{Dias2005}, which were determined considering $\Omega_p = 25 \pm 1$ km\,s$^{-1}$\,kpc$^{-1}$, for $R_{0}=7$ kpc and $V_{0}=200$ km\,s$^{-1}$.

\begin{table}
  \caption[]{$\Omega_p$ (in km\,s$^{-1}$\,kpc$^{-1}$) obtained with Galactic rotation curves from the equations \ref{eq:Vrot10} (plane curve) and \ref{eq:Vrot12} (curve with a dip centered at 9.2 kpc) adopting $R_{0}=8.3$ kpc and $V_{0}=240$ km\,s$^{-1}$. The method used is the linear fit to $\frac{\Delta\theta}{\Delta T}$. 
  In the first column are given the functions fitted used to the location of the zero-age arms: our polynomials (poly) and spiral arms log-periodic functions (spiral) from \citet{Reid2014ApJ...783..130R}.
  The value of all arms implies that the corotation radius is close to the solar Galactic orbit ($R_c/R_0 = 1.02\pm0.07$). See the text for details.}   
\label{tab:omegap}
\begin{center}
\begin{tabular}{llcc}
\hline 
      &arm   & Eq. \ref{eq:Vrot10}    &    Eq. \ref{eq:Vrot12} \\ 
poly  &all   & $28.1\pm1.2$           &    $28.1\pm1.3$         \\     
      &Pers  & $28.2\pm1.7$           &    $28.4\pm1.7$         \\     
      &Loc   & $28.6\pm1.1$           &    $28.9\pm1.1$         \\     
      &Car   & $27.3\pm3.7$           &    $27.0\pm3.6$         \\     
      \hline % ----- spiral
spiral&all   & $28.4\pm1.7$           &    $28.2\pm1.6$         \\     
      &Pers  & $27.0\pm1.7$           &    $26.1\pm1.2$         \\     
      &Loc   & $27.7\pm1.9$           &    $27.5\pm1.8$         \\     
      &Car   & $20.9\pm3.2$           &    $21.2\pm3.2$         \\     
      \hline % ----- mean
mean  &all   & $28.2\pm2.1$           &    $28.2\pm2.1$         \\     
      &Pers  & $27.6\pm2.4$           &    $27.2\pm2.1$         \\     
      &Loc   & $28.2\pm2.5$           &    $28.2\pm2.1$         \\     
      &Car   & $24.1\pm4.9$           &    $24.1\pm4.9$         \\     
\hline
\end{tabular}
\end{center}
\end{table}

Finaly, the determined values of $\Omega_p$ show that there is 
no statistical distinction between the $\Omega_p$ of the different arms.

\section{Conclusions}
In this work, after exploring the Gaia DR2 data of several hundreds open clusters, we selected a sub-sample of young (age < 50 Myr) objects situated within about 5 kpc from the Sun, and which presented the complete set of kinematic data and highest quality of the isochrone fit to the photometric data.
We used the Gaia DR2 data to establish the stellar astrometric membership which were used to guide the isochrone fit performed by our global optimization tool to determine reliable and precise distance and age of the clusters. 
We used this homogeneous sub-sample completely based on the Gaia DR2 data to determine the spiral pattern rotation speed  and the corotation radius of the Galaxy.  

The method to determine $\Omega_p$ 
does not make use of any model of the Galactic disc, except adopting an observed rotation curve taken from the literature, and assuming the widely accepted idea that the clusters are born in the spiral arms. A straightforward numerical integration of the cluster's orbits is performed to discover the position of the arms at a time T ago, equal to the age of the cluster. 

The data of the spiral arm segments in the solar neighborhood, Sagittarius-Carina, Local, and Perseus,  indicates that all the three arms were producing, during the last 50 Myr, a steady flow of clusters being born at a regular rate, at different distances from the present position of the spiral arm, as shown in Figure \ref{fig:linearfit}. 
There is no indication of any major burst-like periods of star-formation activity within the short time interval that we sampled. 

In this study using a homogeneous data of open clusters, we determined $\Omega_p$ of the Galaxy to be $28.2 \pm 2.1$ km\,s$^{-1}$\,kpc$^{-1}$, which implies that the corotation radius is located close to the solar Galactic orbit ($R_c/R_0=1.02\pm0.07$).

This result is in agreement with the classical interpretation and observations of the spiral structures, which place corotation about midway between the inner and outer Lindblad resonances \citep{1998ApJ...502..582C}, and is also consistent  with previous results of our group \citep{Dias2005}.

The fact that corotation, the strongest resonance of the Galactic disk, is very close to the radius of the solar orbit, is essential for the understanding of the dynamics of the solar neighborhood, and is rich of consequences. This indicates that the Sun itself is probably trapped in the resonance and gives a hint on the dynamical origin of the Local arm \citep{lepine2017}. It explains the main aspects of the moving groups \citep{Michtchenko2018ApJ...863L..37M} and the step in the metallicity gradient observed by means of the open clusters \citep{Lepine2011} probably associated with the high metallicity of trapped stars.

The presented results show that the Perseus, Local and Sagittarius-Carina arms rotate with the same angular velocity, within the errors of measurements. This means that there is no observational support for theories that consider arms with different velocities. 

Finally, in our opinion the results should be more accurate with the Gaia final data. By improving our isochrone fit code, we may increase the sample of young open cluster with more precise distances and ages, and extend the age range.

\section*{Acknowledgements}
We thank the referee Dr. Mark Reid for his valuable suggestions which improved the text. 
W.S.Dias acknowledges the S\~ao Paulo State Agency
FAPESP (fellowship 2013/01115-6). H. Monteiro would like to thank
FAPEMIG grants APQ-02030-10 and CEX-PPM-00235-12.
This research was performed using the facilities of the Laborat\'orio de Astrof\'isica Computacional da Universidade Federal de Itajub\'a (LAC-UNIFEI).
This work has made use of data from the European Space Agency (ESA) mission Gaia (http://www.cosmos.esa.int/gaia), processed
by the Gaia Data Processing and Analysis Consortium (DPAC,
http://www.cosmos.esa.int/web/gaia/dpac/consortium). We employed catalogues from CDS/Simbad (Strasbourg)
and Digitized Sky Survey images from the Space Telescope Science
Institute (US Government grant NAG W-2166).


\bibliographystyle{mnras}
\bibliography{refs} % if your bibtex file is called example.bib

%%%%%%%%%%%%%%%%% APPENDICES %%%%%%%%%%%%%%%%%%%%%
%\pagestyle{fancy}
%\fancyhf{}

%\clearpage

\appendix
\setcounter{table}{0}
\renewcommand{\thetable}{A\arabic{table}}

\onecolumn

\begin{longtable}{lcccccccccccc}
\caption[]{Results of mean astrometric parameters obtained using the Gaia DR2 stellar proper motion and parallaxes.  
The meaning of the
symbols are as follows:
$N_{c}$ is the number of cluster stars;
R is the radius (in arcmin) used for each cluster to extract the Gaia DR2 data, centered on the coordinates of the cluster obtained from visual inspection. 
$\varpi$ is the mean parallax of the cluster and $\sigma \varpi$ is the dispersion of the mean parallax.
$\mu_{\alpha}cos{\delta}$ and $\mu_{\delta}$ are the  proper motion components in mas yr$^{-1}$;
$\sigma$ is the dispersion of cluster stars' proper motions;
RV and $\sigma RV$ are the mean and 1$\sigma$ dispersion radial velocity obtained for the cluster.}
\label{tab:astrometric}
%\begin{center}
%\begin{tabular}{lccccccc|ccccc}
\\ \hline
name     &   $\alpha$     &     $\delta$       &  $N_{c}$   &  R &  $\varpi$  &  $\sigma \varpi$  &  $\mu_{\alpha}cos{\delta}$   &  $\sigma\mu_{\alpha}cos{\delta}$  &   $\mu_{\delta}$   &  $\sigma \mu_{\delta}$ &   RV   &   $\sigma RV$ \\
         Alessi 20 &  00  10  22&   +58  44  31&     124&  7.46 &  2.311 &  0.068 &    8.195 &   0.027 &   -2.341&    0.028&  -4.314 &  3.034\\
           NGC 146 &  00  32  58&   +63  20  03&     197&  3.32 &  0.311 &  0.026 &   -2.836 &   0.010 &   -0.454&    0.007&  17.840 & 17.630\\
           King 16 &  00  43  45&   +64  11  08&     551&  6.93 &  0.314 &  0.040 &   -2.654 &   0.040 &   -0.409&    0.013& -42.961 &  0.919\\
        Berkeley 4 &  00  45  01&   +64  23  05&     238&  3.34 &  0.284 &  0.044 &   -2.421 &   0.013 &   -0.269&    0.009& -53.130 &  1.190\\
           NGC 366 &  01  06  26&   +62  13  48&     156&  2.09 &  0.328 &  0.050 &   -2.060 &   0.011 &   -0.414&    0.009&  83.140 & 15.820\\
           NGC 457 &  01  19  35&   +58  17  12&    1079&  6.35 &  0.297 &  0.050 &   -1.578 &   0.006 &   -0.651&    0.009&  79.484 &  2.578\\
           NGC 581 &  01  33  23&   +60  39  00&     237&  3.18 &  0.370 &  0.032 &   -1.382 &   0.004 &   -0.504&    0.010& -45.330 &  0.320\\
          FSR 0551 &  01  39  45&   +64  42  41&      52&  3.95 &  1.063 &  0.052 &    0.112 &   0.022 &   -1.422&    0.021& -23.490 &  5.290\\
           NGC 659 &  01  44  24&   +60  40  24&     272&  3.17 &  0.272 &  0.050 &   -0.799 &   0.008 &   -0.276&    0.010&  77.810 & 17.610\\
          Riddle 4 &  02  07  23&   +60  15  25&     117&  2.72 &  0.337 &  0.050 &   -0.761 &   0.006 &   -0.523&    0.007& -30.388 &  0.407\\
           NGC 884 &  02  22  23&   +57  07  33&    1253&  7.24 &  0.395 &  0.038 &   -0.607 &   0.007 &   -1.049&    0.009& -43.525 &  0.284\\
           IC 1805 &  02  32  50&   +61  38  16&     136&  5.03 &  0.449 &  0.039 &   -0.702 &   0.015 &   -0.669&    0.018& -40.930 &  3.990\\
            ASCC 9 &  02  46  55&   +57  43  48&     525&  8.33 &  0.391 &  0.034 &    0.156 &   0.012 &   -1.086&    0.010& -72.118 &  2.037\\
          Stock 23 &  03  16  11&   +60  06  56&      62& 10.43 &  1.620 &  0.003 &   -4.286 &   0.005 &   -0.921&    0.013& -16.298 &  6.274\\
        Czernik 15 &  03  23  12&   +52  15  00&      74&  2.58 &  0.309 &  0.050 &    0.396 &   0.007 &   -1.082&    0.006& -84.820 &  1.800\\
         Juchert 9 &  03  55  22&   +58  23  28&      39&  1.34 &  0.186 &  0.087 &   -0.223 &   0.018 &   -0.033&    0.023& -34.630 &  0.520\\
          NGC 1502 &  04  07  49&   +62  19  55&     154&  3.28 &  0.916 &  0.054 &   -0.571 &   0.016 &   -0.848&    0.018& -14.760 &  6.480\\
           ASCC 19 &  05  27  56&   -01  59  13&     188& 25.37 &  2.768 &  0.089 &    1.152 &   0.019 &   -1.234&    0.018&  23.574 &  2.129\\
           ASCC 21 &  05  28  43&   +03  31  37&     131& 17.45 &  2.866 &  0.131 &    1.404 &   0.026 &   -0.632&    0.022&  16.034 &  3.808\\
      Collinder 69 &  05  35  10&   +09  48  47&     669& 37.42 &  2.462 &  0.124 &    1.194 &   0.021 &   -2.118&    0.017&  27.727 &  3.754\\
          NGC 1980 &  05  35  24&   -05  54  54&     122&  7.89 &  2.585 &  0.050 &    1.230 &   0.017 &    0.529&    0.025&  25.264 &  7.055\\
          FSR 0850 &  05  45  15&   +24  45  13&      49&  2.12 &  0.456 &  0.050 &    1.286 &   0.017 &   -2.518&    0.010&  19.370 &  1.350\\
      Collinder 95 &  06  31  09&   +09  53  38&     144&  6.77 &  1.458 &  0.117 &   -2.263 &   0.028 &   -5.159&    0.026&   8.726 &  0.595\\
          NGC 2244 &  06  32  11&   +04  54  50&     623&  9.02 &  0.617 &  0.127 &   -1.598 &   0.019 &    0.179&    0.017& 102.430 & 17.380\\
     Collinder 107 &  06  36  49&   +04  58  19&     159& 16.89 &  0.617 &  0.067 &   -1.325 &   0.015 &    0.659&    0.013& 131.600 &  7.420\\
          NGC 2264 &  06  40  58&   +09  53  42&     667& 13.27 &  1.355 &  0.050 &   -1.796 &   0.040 &   -3.651&    0.011&  19.006 &  4.567\\
        vdBergh 92 &  07  03  54&   -11  32  00&      21&  2.89 &  0.860 &  0.024 &   -4.469 &   0.013 &    1.461&    0.010&  25.600 & 11.310\\
     Collinder 132 &  07  13  56&   -30  45  29&      99& 61.20 &  1.501 &  0.087 &   -4.140 &   0.019 &    3.732&    0.024&  27.339 &  3.912\\
     Collinder 135 &  07  17  27&   -37  02  39&     352& 46.46 &  3.278 &  0.117 &   -9.975 &   0.027 &    6.157&    0.025&  16.719 &  2.325\\
          NGC 2362 &  07  18  41&   +24  57  14&     165&  2.74 &  0.742 &  0.074 &   -2.791 &   0.014 &    2.953&    0.018&  28.860 &  5.650\\
       Ruprecht 18 &  07  24  39&   -26  13  00&     203&  3.84 &  0.372 &  0.029 &   -0.495 &   0.006 &    1.159&    0.007&  31.403 &  0.373\\
     Collinder 140 &  07  27  32&   -31  57  58&     150& 22.07 &  2.594 &  0.105 &   -8.074 &   0.022 &    4.789&    0.024&  16.018 &  3.503\\
          NGC 2414 &  07  33  12&   -15  27  12&     174&  2.72 &  0.171 &  0.050 &   -1.393 &   0.006 &    1.460&    0.014& 194.970 &  3.020\\
          NGC 2439 &  07  40  46&   -31  41  38&     551&  3.15 &  0.232 &  0.047 &   -2.283 &   0.004 &    3.159&    0.005&  78.975 &  1.239\\
        Haffner 13 &  07  40  50&   -30  04  23&     210& 13.58 &  1.742 &  0.074 &   -6.184 &   0.018 &    5.879&    0.015&  33.488 &  7.284\\
         NGC 2451B &  07  44  31&   -37  57  14&     298& 23.83 &  2.719 &  0.095 &   -9.671 &   0.034 &    4.702&    0.025&  22.791 &  2.739\\
        Haffner 15 &  07  45  31&   -32  50  46&     175&  1.80 &  0.238 &  0.051 &   -2.154 &   0.009 &    3.289&    0.011&  52.950 &  0.420\\
          NGC 2453 &  07  47  35&   -27  11  42&     271&  2.44 &  0.224 &  0.048 &   -2.346 &   0.012 &    3.430&    0.020&  64.721 &  0.333\\
       Ruprecht 44 &  07  58  51&   -28  35  00&     468&  4.87 &  0.172 &  0.034 &   -2.338 &   0.018 &    2.843&    0.006&  64.017 &  0.467\\
           Pozzo 1 &  08  09  30&   -47  20  06&     390& 22.00 &  2.853 &  0.102 &   -6.516 &   0.015 &    9.530&    0.019&  19.189 &  2.993\\
          NGC 2547 &  08  10  09&   -49  12  54&     475&  9.34 &  2.550 &  0.048 &   -8.611 &   0.051 &    4.257&    0.041&  12.139 &  2.064\\
           IC 2395 &  08  42  30&   -48  06  48&     320&  6.41 &  1.382 &  0.025 &   -4.385 &   0.017 &    3.199&    0.009&  23.197 &  4.154\\
     Collinder 197 &  08  44  51&   -41  14  00&     154&  6.98 &  1.029 &  0.050 &   -5.728 &   0.051 &    4.026&    0.017&  30.789 &  6.426\\
       Trumpler 10 &  08  47  54&   -42  27  00&     151& 11.51 &  2.263 &  0.050 &  -12.385 &   0.083 &    6.465&    0.043&  20.183 &  3.097\\
         Alessi 43 &  08  50  17&   -41  43  12&     427& 14.04 &  1.036 &  0.024 &   -5.472 &   0.032 &    3.931&    0.029&  67.870 & 15.330\\
           IC 2581 &  10  27  29&   -57  37  00&     198&  3.00 &  0.349 &  0.025 &   -7.254 &   0.020 &    3.614&    0.016&  -4.617 &  0.225\\
          NGC 3293 &  10  35  51&   -58  13  48&     233&  2.86 &  0.379 &  0.025 &   -7.671 &   0.032 &    3.350&    0.019& -13.160 &  0.550\\
             BH 99 &  10  38  13&   -59  10  05&     389& 19.92 &  2.225 &  0.076 &  -14.494 &   0.016 &    0.919&    0.017&  11.617 &  6.317\\
       Trumpler 15 &  10  44  43&   -59  22  08&     154&  1.93 &  0.398 &  0.041 &   -6.209 &   0.019 &    2.016&    0.012&  16.590 &  0.290\\
       Trumpler 18 &  11  11  28&   -60  40  00&     213&  3.32 &  0.633 &  0.028 &   -7.168 &   0.011 &    0.583&    0.018&  -4.600 &  0.290\\
       Ruprecht 94 &  11  30  37&   -63  26  00&     763&  7.82 &  0.369 &  0.038 &   -6.366 &   0.013 &    0.886&    0.020&   1.633 &  1.097\\
          NGC 3766 &  11  36  14&   -61  36  30&     837&  4.37 &  0.460 &  0.035 &   -6.733 &   0.017 &    1.010&    0.009& -16.489 &  0.269\\
          Stock 14 &  11  43  48&   -62  31  00&      89&  3.78 &  0.391 &  0.009 &   -6.364 &   0.018 &    0.847&    0.021&  -3.980 &  0.360\\
     Collinder 359 &  12  02  24&   +03  15  36&     329& 62.16 &  1.786 &  0.105 &    0.637 &   0.014 &   -8.668&    0.015&  -3.243 &  0.986\\
          Basel 18 &  13  27  44&   -62  18  46&     117&  3.10 &  0.526 &  0.030 &   -5.063 &   0.020 &   -2.081&    0.039&  12.240 &  0.390\\
     Collinder 272 &  13  30  26&   -61  19  00&     466&  4.31 &  0.420 &  0.033 &   -3.462 &   0.012 &   -1.861&    0.019& -48.400 &  1.690\\
           ASCC 79 &  15  18  55&   -60  47  53&     129& 30.15 &  1.176 &  0.067 &   -2.914 &   0.036 &   -4.232&    0.037&  -4.619 &  7.764\\
          NGC 6193 &  16  41  20&   -48  45  48&     130&  4.65 &  0.823 &  0.046 &    1.401 &   0.019 &   -3.967&    0.021& -14.539 &  3.860\\
           Hogg 21 &  16  45  37&   -47  44  00&     234&  2.63 &  0.347 &  0.035 &   -0.960 &   0.012 &   -2.174&    0.008& -41.031 &  0.669\\
          NGC 6216 &  16  49  24&   -44  43  42&     316&  2.83 &  0.362 &  0.029 &   -1.245 &   0.017 &   -2.550&    0.018& -34.870 &  0.400\\
            BH 200 &  16  49  56&   -44  11  00&     108&  2.62 &  0.409 &  0.039 &   -0.112 &   0.000 &   -1.026&    0.006&  -4.960 &  0.650\\
          NGC 6231 &  16  54  10&   -41  49  30&     517&  5.62 &  0.591 &  0.041 &   -0.569 &   0.006 &   -2.164&    0.006& -28.470 & 15.750\\
           ASCC 88 &  17  06  47&   -35  36  00&    1407& 22.93 &  0.890 &  0.027 &    1.254 &   0.013 &   -2.256&    0.018&  -3.904 &  2.149\\
          NGC 6322 &  17  18  30&   -42  56  13&      70&  2.90 &  0.728 &  0.093 &    0.183 &   0.026 &   -2.272&    0.023& -40.338 &  1.076\\
            BH 231 &  17  31  56&   -31  54  36&      96&  2.33 &  0.328 &  0.027 &   -0.851 &   0.011 &   -1.852&    0.006& -11.340 &  0.320\\
       Trumpler 28 &  17  36  55&   -32  28  08&     143&  3.97 &  0.678 &  0.077 &   -0.851 &   0.021 &   -2.824&    0.018& -15.810 &  1.310\\
       Trumpler 33 &  18  24  42&   -19  43  00&      98&  3.03 &  0.699 &  0.040 &    0.458 &   0.010 &    0.511&    0.005&  -4.640 &  0.380\\
          NGC 6664 &  18  36  30&   -08  11  38&     237&  4.49 &  0.468 &  0.059 &   -0.089 &   0.008 &   -2.561&    0.008&  -4.440 &  0.360\\
         Roslund 2 &  19  45  24&   +23  55  00&     337&  6.75 &  0.458 &  0.034 &   -1.789 &   0.018 &   -5.094&    0.020& 107.340 &  7.110\\
           IC 4996 &  20  16  31&   +37  39  19&     143&  2.16 &  0.479 &  0.029 &   -2.652 &   0.020 &   -5.272&    0.013& -34.614 &  2.477\\
       Berkeley 86 &  20  20  24&   +38  42  00&     143&  3.31 &  0.572 &  0.041 &   -3.510 &   0.022 &   -5.497&    0.027&  -3.733 &  0.671\\
        Teutsch 30 &  20  27  43&   +36  04  32&      23&  1.77 &  0.534 &  0.018 &   -3.172 &   0.034 &   -5.682&    0.024&  94.080 & 14.030\\
          NGC 7128 &  21  43  58&   +53  42  54&     180&  1.64 &  0.261 &  0.061 &   -4.026 &   0.013 &   -4.111&    0.011& -54.930 &  0.530\\
  Alessi Teutsch 5 &  22  08  52&   +61  06  11&     158& 15.61 &  1.111 &  0.062 &   -1.895 &   0.032 &   -3.207&    0.018&  -5.040 &  4.330\\
          NGC 7235 &  22  12  25&   +57  16  12&     245&  3.00 &  0.262 &  0.036 &   -3.777 &   0.012 &   -3.066&    0.022& -57.120 &  0.190\\
   Pismis Moreno 1 &  22  18  48&   +63  16  00&      45&  3.38 &  1.057 &  0.029 &   -2.286 &   0.037 &   -2.250&    0.015&   1.650 &  9.840\\
       Berkeley 96 &  22  29  49&   +55  23  47&     172&  2.37 &  0.242 &  0.044 &   -3.511 &   0.013 &   -2.976&    0.018&   1.750 &  0.600\\
          NGC 7380 &  22  47  21&   +58  07  54&     558&  8.40 &  0.304 &  0.050 &   -2.624 &   0.015 &   -2.044&    0.026& -51.118 &  2.610\\
          NGC 7510 &  23  11  04&   +60  34  44&     324&  2.36 &  0.286 &  0.048 &   -3.664 &   0.008 &   -2.193&    0.008&  63.740 &  3.440\\
      Negueruela 1 &  23  47  24&   +63  13  04&      53&  1.48 &  0.289 &  0.059 &   -2.988 &   0.031 &   -1.343&    0.023&  -5.020 &  2.940\\
\hline
%\end{tabular}
%\end{center}
\end{longtable}

\clearpage
\begin{longtable}{lcccccccccc}
\caption[]{Results of fundamental parameters obtained from the isochrone fit. The distances are given in pc and ages in logt.   
}
\label{tab:photometric}
%\begin{center}
%\begin{tabular}{lccccccc|ccc}
\\ \hline
           name     &  dist  & $\sigma dist$ & age   & $\sigma age$ &  E(B-V)   & $\sigma E(B-V)$ &  Rv   &  $\sigma Rv$  &  Z     & $\sigma Z$     \\         
         Alessi 20  &         436 &    12   & 6.975 & 0.087 &   0.26 &  0.05 &   3.27 &  0.36 &  0.010 & 0.001\\
           NGC 146  &        2727 &   400   & 7.510 & 0.147 &   0.43 &  0.02 &   3.33 &  0.16 &  0.005 & 0.009\\
           King 16  &        2494 &   101   & 7.203 & 0.056 &   0.67 &  0.04 &   3.02 &  0.17 &  0.002 & 0.001\\
        Berkeley 4  &        2744 &   395   & 7.181 & 0.075 &   0.65 &  0.04 &   2.80 &  0.20 &  0.005 & 0.007\\
           NGC 366  &        2667 &   254   & 7.302 & 0.087 &   0.94 &  0.07 &   2.91 &  0.20 &  0.003 & 0.004\\
           NGC 457  &        2271 &   233   & 7.431 & 0.101 &   0.37 &  0.04 &   3.39 &  0.35 &  0.003 & 0.004\\
           NGC 581  &        2061 &   258   & 7.336 & 0.069 &   0.34 &  0.04 &   3.38 &  0.36 &  0.003 & 0.005\\
          FSR 0551  &         829 &     3   & 7.207 & 0.048 &   0.45 &  0.02 &   2.72 &  0.20 &  0.006 & 0.003\\
           NGC 659  &        3018 &   260   & 7.436 & 0.077 &   0.55 &  0.02 &   3.13 &  0.17 &  0.009 & 0.005\\
          Riddle 4  &        2270 &   371   & 7.328 & 0.099 &   0.74 &  0.06 &   3.39 &  0.26 &  0.003 & 0.006\\
           NGC 884  &        1989 &   222   & 7.232 & 0.053 &   0.41 &  0.01 &   3.33 &  0.10 &  0.004 & 0.004\\
           IC 1805  &        2158 &   211   & 6.878 & 0.055 &   0.73 &  0.08 &   2.56 &  0.35 &  0.019 & 0.004\\
            ASCC 9  &        2047 &   105   & 7.268 & 0.035 &   0.81 &  0.06 &   2.56 &  0.17 &  0.002 & 0.001\\
          Stock 23  &         726 &    32   & 7.641 & 0.084 &   0.33 &  0.02 &   2.45 &  0.24 &  0.030 & 0.009\\
        Czernik 15  &        2605 &   295   & 7.491 & 0.071 &   0.89 &  0.07 &   2.30 &  0.22 &  0.003 & 0.006\\
         Juchert 9  &        7632 &  1461   & 6.896 & 0.171 &   0.74 &  0.09 &   2.86 &  0.48 &  0.021 & 0.005\\
          NGC 1502  &        1243 &    66   & 7.010 & 0.029 &   0.73 &  0.07 &   2.40 &  0.26 &  0.025 & 0.005\\
           ASCC 19  &         354 &    12   & 7.141 & 0.053 &   0.07 &  0.02 &   2.70 &  0.24 &  0.021 & 0.004\\
           ASCC 21  &         353 &    12   & 7.090 & 0.034 &   0.05 &  0.01 &   3.04 &  0.04 &  0.018 & 0.002\\
      Collinder 69  &         401 &     9   & 6.965 & 0.036 &   0.10 &  0.01 &   3.02 &  0.03 &  0.014 & 0.002\\
          NGC 1980  &         297 &    30   & 6.932 & 0.031 &   0.03 &  0.01 &   2.96 &  0.22 &  0.006 & 0.002\\
          FSR 0850  &        1861 &   166   & 7.565 & 0.233 &   0.48 &  0.05 &   3.37 &  0.30 &  0.002 & 0.002\\
      Collinder 95  &         630 &    51   & 6.715 & 0.143 &   0.19 &  0.12 &   2.16 &  0.26 &  0.053 & 0.008\\
          NGC 2244  &        1304 &    29   & 6.904 & 0.120 &   0.44 &  0.01 &   2.63 &  0.09 &  0.005 & 0.003\\
     Collinder 107  &        1328 &    26   & 7.146 & 0.026 &   0.41 &  0.03 &   2.64 &  0.20 &  0.004 & 0.001\\
          NGC 2264  &         600 &    14   & 6.798 & 0.086 &   0.08 &  0.01 &   3.07 &  0.13 &  0.005 & 0.002\\
        vdBergh 92  &        1085 &   124   & 6.852 & 0.132 &   0.30 &  0.04 &   2.23 &  0.20 &  0.034 & 0.003\\
     Collinder 132  &         660 &    37   & 7.668 & 0.035 &   0.05 &  0.01 &   3.15 &  0.15 &  0.015 & 0.005\\
     Collinder 135  &         310 &     8   & 7.575 & 0.029 &   0.07 &  0.01 &   2.87 &  0.17 &  0.018 & 0.002\\
          NGC 2362  &        1285 &   123   & 6.872 & 0.120 &   0.17 &  0.03 &   2.25 &  0.26 &  0.005 & 0.004\\
       Ruprecht 18  &        2317 &   181   & 7.676 & 0.155 &   0.57 &  0.04 &   3.15 &  0.23 &  0.004 & 0.002\\
     Collinder 140  &         398 &     7   & 7.654 & 0.030 &   0.03 &  0.01 &   2.99 &  0.23 &  0.019 & 0.003\\
          NGC 2414  &        4467 &   449   & 7.233 & 0.123 &   0.53 &  0.07 &   2.55 &  0.30 &  0.007 & 0.004\\
          NGC 2439  &        2972 &   104   & 7.463 & 0.057 &   0.33 &  0.03 &   3.14 &  0.20 &  0.002 & 0.000\\
        Haffner 13  &         529 &    14   & 7.560 & 0.043 &   0.07 &  0.00 &   2.04 &  0.07 &  0.013 & 0.002\\
         NGC 2451B  &         370 &     9   & 7.557 & 0.028 &   0.09 &  0.01 &   2.90 &  0.18 &  0.018 & 0.003\\
        Haffner 15  &        4686 &   144   & 7.141 & 0.011 &   0.81 &  0.04 &   3.11 &  0.15 &  0.023 & 0.001\\
          NGC 2453  &        3556 &    96   & 7.321 & 0.107 &   0.38 &  0.02 &   3.48 &  0.17 &  0.002 & 0.000\\
       Ruprecht 44  &        4339 &   308   & 7.358 & 0.095 &   0.44 &  0.03 &   3.37 &  0.20 &  0.008 & 0.002\\
           Pozzo 1  &         341 &     3   & 7.073 & 0.030 &   0.08 &  0.02 &   2.24 &  0.10 &  0.011 & 0.001\\
          NGC 2547  &         377 &    23   & 7.564 & 0.032 &   0.08 &  0.01 &   2.88 &  0.31 &  0.016 & 0.005\\
           IC 2395  &         817 &    87   & 7.051 & 0.099 &   0.15 &  0.02 &   2.78 &  0.20 &  0.010 & 0.007\\
     Collinder 197  &         594 &    25   & 7.111 & 0.039 &   0.47 &  0.05 &   3.24 &  0.29 &  0.003 & 0.001\\
       Trumpler 10  &         442 &    10   & 7.627 & 0.016 &   0.05 &  0.01 &   2.71 &  0.15 &  0.019 & 0.002\\
         Alessi 43  &         649 &    48   & 6.940 & 0.050 &   0.20 &  0.02 &   3.15 &  0.22 &  0.003 & 0.001\\
           IC 2581  &        2272 &   220   & 7.151 & 0.031 &   0.35 &  0.02 &   3.03 &  0.16 &  0.007 & 0.006\\
          NGC 3293  &        1769 &   120   & 7.137 & 0.039 &   0.27 &  0.03 &   2.78 &  0.23 &  0.002 & 0.002\\
             BH 99  &         439 &    14   & 7.640 & 0.056 &   0.06 &  0.01 &   2.64 &  0.32 &  0.017 & 0.002\\
       Trumpler 15  &        3691 &   195   & 6.681 & 0.028 &   0.47 &  0.05 &   2.83 &  0.31 &  0.042 & 0.004\\
       Trumpler 18  &        1553 &   112   & 7.679 & 0.083 &   0.23 &  0.02 &   3.31 &  0.20 &  0.017 & 0.006\\
       Ruprecht 94  &        1798 &    96   & 7.135 & 0.023 &   0.32 &  0.05 &   2.83 &  0.28 &  0.003 & 0.001\\
          NGC 3766  &        1802 &   191   & 7.400 & 0.047 &   0.23 &  0.01 &   2.74 &  0.18 &  0.009 & 0.004\\
          Stock 14  &        1989 &   256   & 7.211 & 0.045 &   0.28 &  0.02 &   2.42 &  0.08 &  0.008 & 0.005\\
     Collinder 359  &         604 &    13   & 7.573 & 0.064 &   0.16 &  0.01 &   3.07 &  0.14 &  0.027 & 0.004\\
          Basel 18  &        1869 &   281   & 7.242 & 0.038 &   0.29 &  0.01 &   2.96 &  0.13 &  0.016 & 0.010\\
     Collinder 272  &        1652 &   195   & 7.353 & 0.072 &   0.51 &  0.06 &   2.42 &  0.31 &  0.003 & 0.005\\
           ASCC 79  &         813 &    38   & 6.930 & 0.150 &   0.19 &  0.02 &   3.46 &  0.15 &  0.007 & 0.003\\
          NGC 6193  &        1031 &    46   & 6.791 & 0.065 &   0.51 &  0.05 &   2.26 &  0.24 &  0.023 & 0.004\\
           Hogg 21  &        2637 &   208   & 7.621 & 0.172 &   0.50 &  0.04 &   3.04 &  0.25 &  0.023 & 0.008\\
          NGC 6216  &        2407 &   413   & 7.626 & 0.167 &   0.84 &  0.11 &   2.47 &  0.34 &  0.015 & 0.010\\
            BH 200  &        1974 &    89   & 7.632 & 0.136 &   0.69 &  0.05 &   3.01 &  0.28 &  0.003 & 0.001\\
          NGC 6231  &         989 &    72   & 7.095 & 0.035 &   0.50 &  0.03 &   2.29 &  0.18 &  0.003 & 0.001\\
           ASCC 88  &         991 &    63   & 7.159 & 0.068 &   0.43 &  0.05 &   3.62 &  0.49 &  0.010 & 0.004\\
          NGC 6322  &        1501 &   136   & 6.968 & 0.097 &   0.59 &  0.10 &   2.92 &  0.44 &  0.017 & 0.007\\
            BH 231  &        2643 &   109   & 7.075 & 0.067 &   0.93 &  0.07 &   2.94 &  0.20 &  0.013 & 0.003\\
       Trumpler 28  &        1380 &    78   & 7.503 & 0.035 &   0.57 &  0.02 &   2.94 &  0.08 &  0.017 & 0.006\\
       Trumpler 33  &        1491 &    49   & 7.435 & 0.109 &   0.42 &  0.04 &   2.98 &  0.19 &  0.015 & 0.003\\
          NGC 6664  &        1897 &   265   & 7.379 & 0.159 &   0.70 &  0.02 &   2.90 &  0.15 &  0.002 & 0.007\\
         Roslund 2  &        1371 &    78   & 7.100 & 0.032 &   0.89 &  0.05 &   2.48 &  0.16 &  0.002 & 0.000\\
           IC 4996  &        2124 &   399   & 6.999 & 0.063 &   0.55 &  0.05 &   2.64 &  0.23 &  0.012 & 0.009\\
       Berkeley 86  &        1861 &   144   & 7.192 & 0.052 &   0.79 &  0.11 &   2.64 &  0.32 &  0.026 & 0.006\\
        Teutsch 30  &        2492 &   207   & 7.506 & 0.139 &   0.95 &  0.04 &   3.02 &  0.17 &  0.022 & 0.003\\
          NGC 7128  &        4824 &   908   & 7.145 & 0.217 &   0.92 &  0.12 &   2.57 &  0.33 &  0.019 & 0.011\\
  Alessi Teutsch 5  &         804 &     9   & 7.118 & 0.051 &   0.40 &  0.04 &   2.38 &  0.30 &  0.006 & 0.002\\
          NGC 7235  &        3737 &   145   & 7.309 & 0.061 &   0.82 &  0.06 &   2.55 &  0.19 &  0.010 & 0.003\\
   Pismis Moreno 1  &         817 &   167   & 7.082 & 0.227 &   0.62 &  0.08 &   2.46 &  0.24 &  0.007 & 0.003\\
       Berkeley 96  &        3196 &   404   & 7.268 & 0.044 &   0.52 &  0.05 &   2.73 &  0.22 &  0.006 & 0.005\\
          NGC 7380  &        1810 &   497   & 7.352 & 0.039 &   0.52 &  0.03 &   2.85 &  0.07 &  0.002 & 0.010\\
          NGC 7510  &        3497 &    99   & 7.100 & 0.040 &   0.86 &  0.02 &   2.78 &  0.07 &  0.015 & 0.002\\
      Negueruela 1  &        5691 &   370   & 6.822 & 0.035 &   0.92 &  0.08 &   3.00 &  0.26 &  0.036 & 0.004\\
\hline
%\end{tabular}
%\end{center}
\end{longtable}

\clearpage
\begin{longtable}{lcccccccc}
\caption[]{Present-day and birthplace positions of the open clusters and their respective errors. The x-axis pointing to the Galactic rotation direction and
the y-axis positive points towards the Galactic anti-center. The Sun is situated at (0,8.3) kpc position. See the text for details }
\label{tab:galactic}
%\begin{center}
%\begin{tabular}{lcccccccc}
\\ \hline
nome             &    X        & $\sigma X$&    Y   & $\sigma Y$&   $X_0$  &$\sigma X_0$ & $Y_0$    & $\sigma Y_0$   \\
    Alessi 20    &   0.386     &    0.011&     8.502&     0.006 &  -1.929  &        0.030&    8.177 &   0.010        \\
      NGC 146    &   2.341     &    0.343&     9.699&     0.205 &  -6.905  &        0.528&    7.774 &   0.482        \\
      King 16    &   2.112     &    0.086&     9.625&     0.054 &  -1.749  &        0.083&    9.871 &   0.067        \\
   Berkeley 4    &   2.320     &    0.334&     9.764&     0.211 &  -1.254  &        0.321&   10.131 &   0.264        \\
      NGC 366    &   2.193     &    0.209&     9.818&     0.145 &  -4.683  &        0.354&    8.293 &   0.221        \\
      NGC 457    &   1.817     &    0.186&     9.651&     0.139 &  -6.853  &        0.169&    6.809 &   0.170        \\
      NGC 581    &   1.623     &    0.203&     9.569&     0.159 &  -3.351  &        0.212&    9.250 &   0.160        \\
     FSR 0551    &   0.652     &    0.002&     8.810&     0.002 &  -3.130  &        0.076&    8.341 &   0.039        \\
      NGC 659    &   2.332     &    0.201&    10.214&     0.165 &  -6.395  &        0.377&    7.226 &   0.335        \\
     Riddle 4    &   1.680     &    0.275&     9.825&     0.249 &  -3.415  &        0.305&    9.290 &   0.225        \\
      NGC 884    &   1.403     &    0.157&     9.704&     0.157 &  -2.495  &        0.180&    9.605 &   0.139        \\
      IC 1805    &   1.533     &    0.150&     9.819&     0.148 &  -0.239  &        0.149&    9.982 &   0.150        \\
       ASCC 9    &   1.376     &    0.071&     9.815&     0.078 &  -2.419  &        0.088&    9.973 &   0.069        \\
     Stock 23    &   0.464     &    0.020&     8.858&     0.025 &  -8.764  &        0.126&    4.606 &   0.229        \\
   Czernik 15    &   1.486     &    0.168&    10.432&     0.241 &  -4.548  &        0.254&   10.290 &   0.209        \\
    Juchert 9    &   4.355     &    0.834&    14.548&     1.196 &  +2.355  &        0.661&   14.794 &   1.296        \\
     NGC 1502    &   0.730     &    0.039&     9.292&     0.053 &  -1.820  &        0.066&    9.186 &   0.063        \\
      ASCC 19    &  -0.141     &    0.005&     8.603&     0.010 &  -3.448  &        0.013&    7.827 &   0.030        \\
      ASCC 21    &  -0.115     &    0.004&     8.618&     0.011 &  -3.131  &        0.018&    8.058 &   0.046        \\
 Collinder 69    &  -0.103     &    0.002&     8.679&     0.008 &  -2.348  &        0.010&    8.254 &   0.035        \\
     NGC 1980    &  -0.138     &    0.014&     8.543&     0.025 &  -2.220  &        0.022&    8.198 &   0.060        \\
     FSR 0850    &  -0.115     &    0.010&    10.156&     0.166 &  -7.501  &        0.161&    6.233 &   0.173        \\
 Collinder 95    &  -0.234     &    0.019&     8.885&     0.047 &  -1.500  &        0.011&    8.832 &   0.051        \\
     NGC 2244    &  -0.579     &    0.013&     9.468&     0.026 &  -2.292  &        0.046&    8.576 &   0.138        \\
Collinder 107    &  -0.600     &    0.012&     9.485&     0.023 &  -3.347  &        0.048&    7.277 &   0.104        \\
     NGC 2264    &  -0.231     &    0.005&     8.853&     0.013 &  -1.755  &        0.009&    8.692 &   0.032        \\
   vdBergh 92    &  -0.761     &    0.087&     9.072&     0.088 &  -2.541  &        0.083&    8.773 &   0.120        \\
Collinder 132    &  -0.580     &    0.033&     8.597&     0.017 &  -8.375  &        0.217&    1.557 &   0.102        \\
Collinder 135    &  -0.284     &    0.007&     8.409&     0.003 &  -7.765  &        0.086&    3.637 &   0.042        \\
     NGC 2362    &  -1.087     &    0.104&     8.974&     0.065 &  -2.891  &        0.092&    8.614 &   0.091        \\
  Ruprecht 18    &  -1.997     &    0.156&     9.459&     0.091 &  -9.891  &        0.240&    3.297 &   0.120        \\
Collinder 140    &  -0.357     &    0.006&     8.467&     0.003 &  -8.551  &        0.185&    2.017 &   0.081        \\
     NGC 2414    &  -3.490     &    0.351&    11.084&     0.280 &  -5.381  &        0.289&    7.878 &   0.377        \\
     NGC 2439    &  -2.716     &    0.095&     9.484&     0.041 &  -7.508  &        0.099&    5.178 &   0.078        \\
   Haffner 13    &  -0.479     &    0.013&     8.523&     0.006 &  -7.267  &        0.272&    3.515 &   0.136        \\
    NGC 2451B    &  -0.350     &    0.009&     8.412&     0.003 &  -7.377  &        0.095&    3.854 &   0.050        \\
   Haffner 15    &  -4.332     &    0.133&    10.055&     0.054 &  -7.439  &        0.100&    8.355 &   0.102        \\
     NGC 2453    &  -3.176     &    0.086&     9.899&     0.043 &  -7.459  &        0.075&    7.089 &   0.073        \\
  Ruprecht 44    &  -3.955     &    0.281&    10.083&     0.127 &  -8.477  &        0.251&    6.963 &   0.227        \\
      Pozzo 1    &  -0.335     &    0.003&     8.342&     0.000 &  -3.117  &        0.029&    7.724 &   0.020        \\
     NGC 2547    &  -0.371     &    0.023&     8.335&     0.002 &  -7.679  &        0.067&    3.890 &   0.043        \\
      IC 2395    &  -0.814     &    0.087&     8.349&     0.005 &  -3.381  &        0.086&    7.788 &   0.045        \\
Collinder 197    &  -0.587     &    0.025&     8.388&     0.004 &  -3.446  &        0.069&    7.642 &   0.052        \\
  Trumpler 10    &  -0.439     &    0.010&     8.354&     0.001 &  -7.817  &        0.137&    1.979 &   0.042        \\
    Alessi 43    &  -0.643     &    0.048&     8.383&     0.006 &  -2.275  &        0.131&    7.959 &   0.065        \\
      IC 2581    &  -2.199     &    0.213&     7.727&     0.055 &  -5.286  &        0.173&    5.974 &   0.154        \\
     NGC 3293    &  -1.702     &    0.115&     7.817&     0.033 &  -4.900  &        0.100&    6.317 &   0.079        \\
        BH 99    &  -0.421     &    0.013&     8.175&     0.004 &  -7.726  &        0.232&    1.655 &   0.095        \\
  Trumpler 15    &  -3.522     &    0.186&     7.196&     0.058 &  -4.482  &        0.161&    6.670 &   0.110        \\
  Trumpler 18    &  -1.450     &    0.105&     7.743&     0.040 &  -8.042  &        0.066&   -0.518 &   0.157        \\
Ruprecht 94      &  -1.641     &    0.088&     7.568&     0.039 &  -4.597  &        0.076&    6.358 &   0.057        \\
     NGC 3766    &  -1.644     &    0.174&     7.562&     0.078 &  -6.791  &        0.123&    4.183 &   0.181        \\
     Stock 14    &  -1.798     &    0.231&     7.450&     0.109 &  -5.226  &        0.188&    5.756 &   0.173        \\
Collinder 359    &   0.297     &    0.006&     7.790&     0.011 &  -6.944  &        0.040&    3.700 &   0.029        \\
     Basel 18    &  -1.490     &    0.224&     7.171&     0.170 &  -4.873  &        0.213&    5.663 &   0.197        \\
Collinder 272    &  -1.308     &    0.154&     7.291&     0.119 &  -6.720  &        0.112&    4.539 &   0.154        \\
      ASCC 79    &  -0.522     &    0.024&     7.678&     0.029 &  -2.586  &        0.063&    7.358 &   0.048        \\
     NGC 6193    &  -0.406     &    0.018&     7.353&     0.042 &  -1.965  &        0.031&    7.152 &   0.042        \\
      Hogg 21    &  -0.989     &    0.078&     5.857&     0.193 &  -5.866  &        0.225&   -1.787 &   0.287        \\
     NGC 6216    &  -0.796     &    0.137&     6.029&     0.390 &  -5.585  &        0.472&   -1.768 &   0.613        \\
       BH 200    &  -0.638     &    0.029&     6.432&     0.084 &  -7.079  &        0.091&    0.154 &   0.114        \\
     NGC 6231    &  -0.283     &    0.021&     7.353&     0.069 &  -3.370  &        0.144&    6.481 &   0.155        \\
      ASCC 88    &  -0.171     &    0.011&     7.326&     0.062 &  -3.675  &        0.050&    6.592 &   0.062        \\
     NGC 6322    &  -0.381     &    0.034&     6.850&     0.131 &  -2.699  &        0.083&    6.165 &   0.113        \\
       BH 231    &  -0.190     &    0.008&     5.664&     0.109 &  -2.819  &        0.068&    4.914 &   0.096        \\
  Trumpler 28    &  -0.096     &    0.005&     6.923&     0.078 &  -6.102  &        0.106&    2.692 &   0.095        \\
  Trumpler 33    &   0.320     &    0.011&     6.846&     0.048 &  -5.866  &        0.051&    3.932 &   0.050        \\
     NGC 6664    &   0.770     &    0.108&     6.566&     0.242 &  -4.432  &        0.255&    4.335 &   0.250        \\
    Roslund 2    &   1.190     &    0.068&     7.619&     0.039 &  -3.024  &        0.086&    8.236 &   0.085        \\
      IC 4996    &   2.055     &    0.386&     7.764&     0.101 &  -0.105  &        0.403&    8.012 &   0.019        \\
  Berkeley 86    &   1.810     &    0.140&     7.870&     0.033 &  -1.997  &        0.146&    8.035 &   0.037        \\
   Teutsch 30    &   2.410     &    0.200&     7.670&     0.052 &  -7.882  &        0.355&    7.827 &   0.354        \\
     NGC 7128    &   4.784     &    0.901&     8.917&     0.116 &  +1.460  &        0.706&   10.534 &   0.584        \\
     
Alessi Teutsch 5 &   0.776     &    0.009&     8.501&     0.002 &  -2.572  &        0.056&    8.233 &   0.018        \\
     NGC 7235    &   3.645     &    0.141&     9.121&     0.032 &  -1.108  &        0.126&   10.181 &   0.103        \\
Pismis Moreno 1  &   0.779     &    0.159&     8.533&     0.048 &  -2.399  &        0.196&    8.298 &   0.069        \\
  Berkeley 96    &   3.103     &    0.392&     9.058&     0.096 &  -2.177  &        0.380&    9.497 &   0.227        \\
     NGC 7380    &   1.729     &    0.475&     8.834&     0.147 &  -3.183  &        0.439&    8.400 &   0.265        \\
     NGC 7510    &   3.267     &    0.092&     9.548&     0.035 &  -1.156  &        0.094&    9.783 &   0.054        \\
 Negueruela 1    &   5.123     &    0.333&    10.776&     0.161 &  +3.124  &        0.246&   11.300 &   0.267        \\
\hline
%\end{tabular}
%\end{center}
\end{longtable}

%%%%%%%%%%%%%%%%%%%%%%%%%%%%%%%%%%%%%%%%%%%%%%%%%%

% Don't change these lines
\bsp	% typesetting comment
\label{lastpage}
\end{document}